\documentclass[sigplan,10pt]{acmart}
\acmConference[PPoPP'26]{ACM SIGPLAN Annual Symposium on Principles and Practice of Parallel Programming}{January 31--February 04, 2026}{Sydney, Australia}
\acmYear{2026}
\acmISBN{} 
\acmDOI{} 
\startPage{1}

\setcopyright{none}

\usepackage{booktabs}   
\usepackage{subcaption} 

\newcommand{\anonymous}[1]{%
  \ifdefined\isanonymoustrue
    \iffalse
      #1
    \else
      [Hidden]
    \fi
  \else
    #1
  \fi
}

\usepackage{acronym}
\usepackage{multicol}
\usepackage{algorithm,algpseudocode}
\usepackage{amsmath}
\usepackage{appendix}
\usepackage{mathtools}
\usepackage{amsfonts}
\usepackage{bm}
\usepackage[font=small,labelfont=bf]{caption}

\usepackage{enumitem} 
\usepackage[c]{esvect}
\usepackage{float}
\usepackage{graphicx}
\graphicspath{{../}{./}}

\usepackage{hyperref}
\hypersetup{pdfauthor=author}

\usepackage{listings}
\usepackage{siunitx}
\usepackage{subcaption}
\usepackage{xspace}

\usepackage[normalem]{ulem}

\usepackage[capitalise]{cleveref}

\usepackage{color}

\usepackage{tikz}
\usetikzlibrary{shapes,arrows}
\usetikzlibrary{shapes.multipart}
\usepackage{adjustbox}
\usepackage{reotex}

\usepackage{overpic}

\definecolor{mygray}{rgb}{0.5,0.5,0.5}
\definecolor{lightgray}{rgb}{0.95,0.95,0.95}

\usepackage{fancyvrb} 
\VerbatimFootnotes 

\definecolor{XcodeComments}{RGB}{00,116,00}
\definecolor{XcodeKeywords}{RGB}{170,13,145}
\definecolor{XcodeStringstyle}{RGB}{21,26,255}
\lstdefinestyle{CppCode}{
  language=[GNU]C++, 
  morekeywords={size_t},
  commentstyle=\color{XcodeComments}\fontseries{b}\fontshape{sl}\selectfont,
  keywordstyle=\color{XcodeKeywords}\fontseries{b}\selectfont,
  stringstyle=\color{XcodeStringstyle}\bfseries,
  emphstyle=\color{black}\bfseries
}

\begin{document}

\title[HiCR]{HiCR, an Abstract Model for Distributed Heterogeneous Programming}         


 \author{Sergio Miguel Martin}
 \author{Luca Terracciano}
\author{Kiril Dichev}
 
\author{Noah Baumann}
 
\author{Orestis Korakitis}
 \affiliation{
   \department{Computing Systems Laboratory} 
   \institution{Huawei Zurich Research Center}            
   \country{Switzerland}
 }
 \authornote{Affiliation at the time of contribution.} 
 
\author{Jiashu Lin}
 \affiliation{
   \institution{HiSilicon Technologies}            
   \country{China}
 }
 
 \author{Albert-Jan Yzelman}
 \affiliation{
   \department{Computing Systems Laboratory} 
   \institution{Huawei Zurich Research Center}            
   \country{Switzerland}
 }
 \email{albertjan.yzelman@huawei.com}          
 \authornote{Corresponding author.}
\fancyhead{}  
\renewcommand\footnotetextcopyrightpermission[1]{} 

\begin{abstract}
We present HiCR, a model to represent the semantics of distributed heterogeneous applications and runtime systems. The model describes a minimal set of abstract operations to enable hardware topology discovery, kernel execution, memory management, communication, and instance management, without prescribing any implementation decisions. The goal of the model is to enable execution in current and future systems without the need for significant refactoring, while also being able to serve any governing parallel programming paradigm. In terms of software abstraction, HiCR is naturally located between distributed heterogeneous systems and runtime systems. We coin the phrase \emph{Runtime Support Layer} for this level of abstraction. We explain how the model's components and operations are realized by a plugin-based approach that takes care of device-specific implementation details, and present examples of HiCR-based applications that operate equally on a diversity of platforms. 
\end{abstract}

\maketitle

\section{Introduction}

Recent advancements in artificial intelligence models present substantial demands to modern computing systems, as they require computational power beyond that of traditional CPUs. In some instances, their memory and compute requirements go beyond what a single device or node can offer. This trend has driven the rise of \textit{distributed heterogeneous systems} as a leading approach for executing AI pipelines. These systems are characterized by, first, the use of \textit{accelerators} such as Graphics and Neural Processing Units (GPUs and NPUs) for higher-dimensional tensor operations, and; second, the use of distributed computing to scale out computational performance and memory capacity. 


Distributed heterogeneous systems induce significant complexities to software development, especially in applications that strive to maximize performance and fully exploit the hardware and interconnect capabilities. On one hand, the use of accelerators often requires the use of vendor-specific interfaces, resulting in tightly coupled implementations. On the other hand, distributed applications must deal with deployment-specific requirements, such as those for cloud platforms or data centers. 

The complexity of handling devices and interconnect-specific technologies undermines the application's portability, as changes to the underlying hardware often necessitate extensive refactoring, even when semantics remain the same. These challenges highlight the need for flexible systems capable of hiding platform-specific implementation details and seamlessly adapt to new hardware and technologies. 

We present HiCR (pronounced as `\emph{hiker}'), a model to represent the semantics of distributed heterogeneous applications and runtime systems. The model exposes a minimal set of operations to enable hardware topology discovery, kernel execution, memory management, communication, and instance management. These operations are not tied to any given technology. Instead, they describe an application in terms of abstract operations that do not prescribe implementation details. As a result, any HiCR-based code will reach its intended result regardless of the system it executes on.

The HiCR model employs a plugin-based approach, delegating to third-party developers the responsibility of translating its semantics into implementation-specific directives. The benefits of this design are twofold. First, code written using HiCR need only be written once to seamlessly execute across a wide range of technologies and platforms. Second, any newly developed plugin automatically becomes available to all HiCR-based code, allowing them to benefit from the added functionality without further modifications.

The contributions of this paper are: (a) the introduction of a \textit{Runtime Support Library} as an intermediate layer between an application and its programming framework or runtime system, and its underlying technologies; (b) the proposal of HiCR, as an abstract model for the operations that this layer should offer; (c) an open-source implementation of the model, and; (d) its functional verification through experiments.

The rest of the paper is as follows: in {\S}\ref{sec:related_work}, we discuss related work and highlight the differences with ours; in {\S}\ref{sec:model} we introduce the HiCR model; in {\S}\ref{sec:implementation}, we describe its implementation and highlight the key components that enables application portability; in {\S}\ref{sec:experiments}, we show empirical results that demonstrate how HiCR-based codes obtain equal results with different technologies without the need of refactoring, and; in {\S}\ref{sec:conclusion} we provide final thoughts and discuss future work.








\section{Related Work}\label{sec:related_work}

This Section discusses similarities and differences of HiCR with, in turn: runtime systems based on metaprogramming (e.g., \textit{pragmas}), tasking runtime systems, skeleton and automatic programming frameworks, as well as (cloud) workflow middleware.

Metaprogramming frameworks expose annotations in the application code to automatically apply transformations during compilation time for the use of specific technologies. \textit{OpenACC}~\cite{openacc}, \textit{OpenArc}~\cite{lee2014} and \textit{HMPP}~\cite{dolbeau2007} detect the functions with annotations and make them available to run on accelerator devices while others, such as \textit{OpenMP}~\cite{openmp, valero-lara2021, ompcluster}, \textit{XcalableMP}~\cite{xcalablemp, xcalableacc}, and \textit{OmpSs-2}~\cite{ompss2, aguilar2022, ayguade2009, elangovan2012} have been extended to support heterogeneous, and in some cases, distributed-memory parallel computing. These programming frameworks provide an advantage when adapting existing code to parallel execution, whereas using HiCR for this purpose may require significant refactoring. However, metaprogramming also introduce complexities in maintaining and extending compiler support for the underlying technologies. Moreover, platform vendors usually provide their own specialized, often closed-source, compiler implementations, thus hindering the extension of these annotations to new system technologies. With HiCR, all operations are explicit and directly handled by the supporting backends, hence preserving compiler compatibility even when extended to new technologies. It also allows for proprietary backends without sacrificing portability.

Task-based programming frameworks organize programs as a collection of tasks and rely on a runtime system to make scheduling and resource allocation decisions. They seamlessly support task execution on heterogeneous devices and across distributed systems by adding specific APIs for such operations. For example, the \textit{StarPU} \cite{augonnet2009} runtime system interfaces directly with MPI and OpenCL or \textit{CUDA}~\cite{cuda} for distributed communication and GPU operations, respectively. For each technology, it adopts corresponding API extensions (\texttt{starpu\_mpi}, \texttt{starpu\_cuda}, and \texttt{star\-pu\_opencl})~\cite{augonnet2012}. The drawback is that either their implementation or API remain fixed to the underlying technologies and are not transferable to any other current or future system. Instead, the HiCR model ensures that applications can seamlessly adapt to future technologies by developing corresponding backends.


Efforts from academia and industry have sought automatic frameworks for distributed and heterogeneous computing that allow expressing portable high-level code. One approach revolves around \emph{skeletons}, mini-programs with prescribed input and output structures that can automatically parallelize and dispatch to accelerators and whose composition describes higher-order functionalities. These include \textit{SkePU}~\cite{enmyren2010}, \textit{Kokkos}~\cite{edwards2012}, \textit{RAJA}~\cite{beckingsale2019}, \textit{AllScale}~\cite{jordan2020}, \textit{SYCL}~\cite{sycl}, and \textit{Data Parallel C++} \cite{dpc++}. In contrast to these approaches, HiCR requires no language extensions nor does it prescribe specific skeletons. PGAS~\cite{padua2011} inspired another set of approaches, leading to frameworks such as \textit{Chapel}~\cite{callahan2004}, \textit{X10}~\cite{charles2005}, and \textit{UPC}~\cite{el-ghazawi2006} -- while a final class of frameworks center around BSP and its automatic PRAM simulation~\cite{valiant1990} which led to frameworks such as MapReduce~\cite{dean2008}, Pregel~\cite{malewicz2010}, and Spark~\cite{zaharia2010}. Applying these approaches to heterogeneous systems is not straightforward; for example, Chapel provides abstractions to define kernels that dispatch to accelerators via ad-hoc extensions such as \texttt{kernel-launch}, and similar for other automatic frameworks~\cite{yuan2016}. In contrast, HiCR provides direct control over system resources and application scheduling decisions without the need for such extensions. 

Distributed memory frameworks such as \textit{COMP Superscalar}~\cite{tejedor2008} enable deployment on cloud infrastructures and dynamic provision of resources in a transparent way to end-users, relieving them from refactoring applications to adapt to different cloud providers. Such functionality is in line with what HiCR ascribes to its \emph{instance manager}. HiCR, however, integrates this functionality into a complete model that also includes memory, compute and communication operations.

Other runtime frameworks~\cite{bosilca2013,gautier2013,kaiser2014,robson2016,zafari2019,gioiosa2020,cardosi2023}
share the similarities and differences with HiCR of the aforementioned approaches.
Others~\cite{rasch2018,thoman2019,c++amp,thrust,hip} allow for general-purpose programming over multiple devices within a single node, but do not handle the aspect of distributed computing.
Of these, IRIS~\cite{kim2021} supports heterogeneous hardware while, similar to HiCR, hiding the device characteristics behind a unique abstract model capable of representing a wide range of hardware. Aside from not supporting for distributed computation, unlike HiCR, IRIS prescribes task-based programming.

The \textit{Open Community Runtime} (OCR)~\cite{mattson2016} provides a common layer for building asynchronous many-tasks runtime systems. OCR focuses on runtime-specific building blocks, like tasks and dependencies, and delegates to each particular implementation (e.g., OCR-Vx~\cite{dokulil2022} for OpenCL) the responsibility for dealing with particular hardware and interconnect technologies. These aims are similar to that of a Runtime Support Layer. However, OCR performs no resource provisioning while implementations must take on dependency graph management and data block management. For distributed-memory implementations the latter are both challenging and prescribe, to significant degree, the design of any task-based programming framework based on it~\cite{dokulil2016}. By contrast, the HiCR API is generic enough to support any programming model, tasking or otherwise, while retaining full freedom as to their implementation.

\section{The HiCR Model}\label{sec:model}

\begin{figure}[htb]
\begin{center}
    \hspace{-40pt} 
\pgfdeclarelayer{background}
\pgfdeclarelayer{foreground}
\pgfsetlayers{background,main,foreground}

\tikzstyle{resource}=[draw, fill=white, font=\scriptsize, text centered, minimum width=20em]
\tikzstyle{device} = [resource, text width=24em, fill=red!20, minimum height=7.0em, rounded corners]
\tikzstyle{manager} = [resource, text width=7em, fill=green!10, minimum height=2.5em, rounded corners]
\tikzstyle{statefulResource}=[resource, text width=7em, fill=yellow!20, draw]
\def\blockdist{2.3}
\def\edgedist{2.5}

\tikzstyle{doc}=[%
draw,
thick,
align=center,
color=black,
shape=document,
minimum width=7mm,
minimum height=13.2mm,
shape=document,
inner sep=2ex,
]

\begin{tikzpicture}
    \node (anchor) [] {};
    
   \path (anchor) node (userApplication) [resource, anchor=north east, minimum width=20em] {User Application};

   \path (userApplication.south east) node (domainSpecificLibrary) [resource, anchor=north east, minimum width=16em] {Domain-Specific Library};

  \path (domainSpecificLibrary.south east) node (runtimeSystem) [resource, anchor=north east, minimum width=12em] {Runtime System};   

  \path (runtimeSystem.south east)+(0.00, -0.25) node (runtimeSupportLibrary) [resource, anchor=north east, fill=black!10] {Runtime Support Library};    

  \path (runtimeSupportLibrary.south)+(0.00, -0.25) node (systemLibraries) [resource, anchor=north] {System Libraries / Plugins};  

  \path (systemLibraries.south) node (deviceDrivers) [resource, anchor=north] {Device Drivers};

  
  \path [draw, ->, line width=0.5mm] ([xshift=-70] userApplication.south -| runtimeSupportLibrary.center) -- ([xshift=-70] runtimeSupportLibrary.north -| runtimeSupportLibrary.center);
  
  \path [draw, ->, line width=0.5mm] ([xshift=-35] domainSpecificLibrary.south -| runtimeSupportLibrary.center) -- ([xshift=-35] runtimeSupportLibrary.north -| runtimeSupportLibrary.center);  
  
  \path [draw, ->, line width=0.5mm] ([xshift=-0] runtimeSystem.south -| runtimeSupportLibrary.center) -- ([xshift=-0] runtimeSupportLibrary.north -| runtimeSupportLibrary.center);
   
  \path [draw, ->, line width=0.5mm] (runtimeSupportLibrary.south -| systemLibraries.center) -- (systemLibraries.north -| systemLibraries.center);

\end{tikzpicture}
\end{center}
\vspace{-10pt}
\caption{A layered view of runtime support for user applications. Any of the layers may invoke functionality of one or more of the layers below it to support runtime operations.}
\label{fig:softwareStack}
\end{figure}
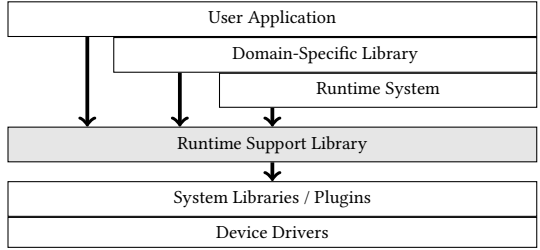

Given the complexity of modern computing systems, applications typically rely on a multitude of third-party and system libraries. \cref{fig:softwareStack} represents the software stack on which a user application may operate at runtime. Any of these layers may rely on functionality provided by one or multiple layers below it. For example, it may suffice to only require access to device drivers (low-level interrupts) or system libraries (e.g., \textit{MPI}~\cite{MPI:forum}). These layers are typically accessed by expert programmers with a deep knowledge of such technologies. Other users may prefer higher productivity at the expense of precise execution control and opt for domain-specific libraries. Intermediate users may prefer a programming model that delegates the complexities of device access and conducting scheduling to an underlying runtime system.

Runtime systems can automatically resolve communication operations and device management directly, encapsulating the required accesses to the corresponding low-level libraries. This approach results in limited portability to other existing or future technologies if the underlying runtime system is not updated accordingly. However, even if it is, updates typically imply runtime system API changes which, in turn, requires refactoring at the application level.

To solve the issue of portability, we propose the concept of a \textit{Runtime Support Layer} to serve as a bridge from application and runtime systems to underlying low-level system libraries. This layer provides an implementation-agnostic API consisting of computation, communication, and system management building blocks. We propose HiCR as a model to describe such building blocks. 


HiCR comprises a minimal set of components and operations to describe the semantics of any code running on any distributed computing system. That is, it assumes the existence of interconnected computing nodes, each comprised of processing units connected to local memory. In addition, the model imposes no semantic prescriptions to the way an application must be programmed or executed. Instead, it can be employed equally well by a programming framework such as a tasking runtime system as well as directly by an application, or indeed by any middleware layer in-between.
By decoupling the runtime support library from the programming model, we thus enable HiCR to be used by highly diverse programming approaches: any runtime system, programming framework, domain-specific library, or user application may delegate all low-level operations to HiCR without the need to consider implementation details. These details are later resolved by device-specific implementations, or \textit{backends}, of the HiCR model. In this way, a wide variety application may be ported to other systems and technologies simply by selecting different sets of backends.

\subsection{Model Description}

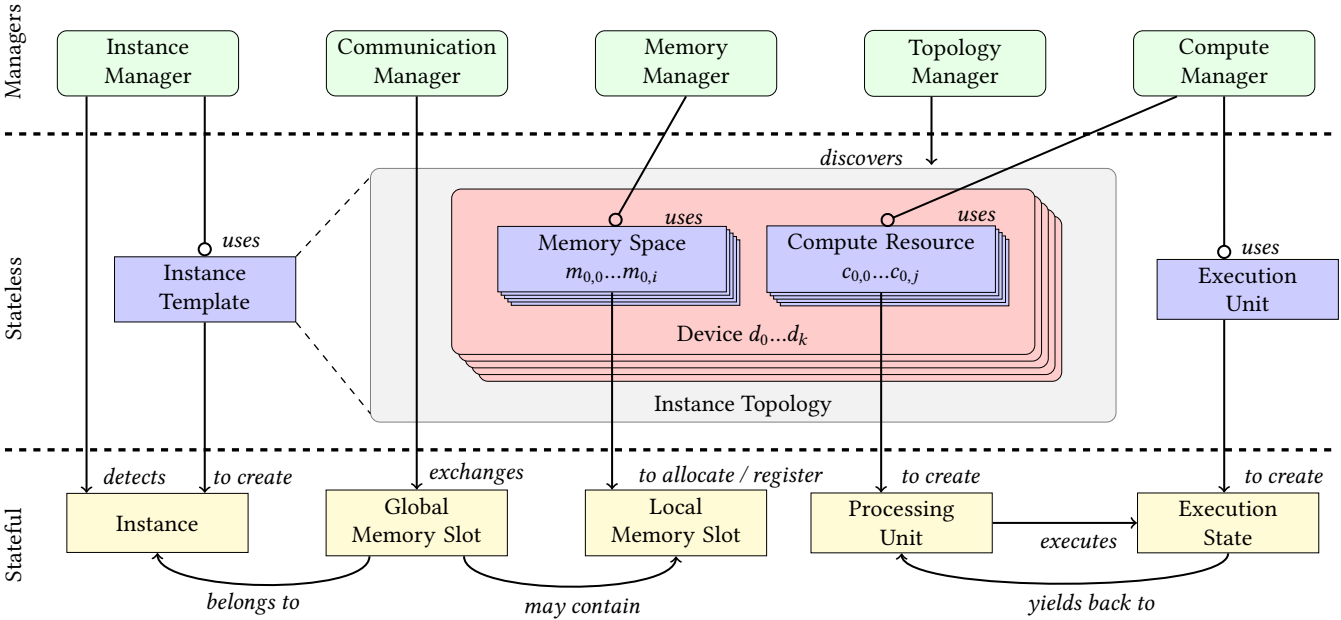
\begin{figure*}[htb]
\begin{center}
    \begin{adjustbox}{width=\textwidth}
    \hspace{-40pt} 
\pgfdeclarelayer{background}
\pgfdeclarelayer{foreground}
\pgfsetlayers{background,main,foreground}

\tikzstyle{resource}=[draw, fill=blue!20, text width=9em, text centered, minimum height=2.5em]
\tikzstyle{device} = [resource, text width=24em, fill=red!20, minimum height=7.0em, rounded corners]
\tikzstyle{manager} = [resource, text width=7em, fill=green!10, minimum height=2.5em, rounded corners]
\tikzstyle{statefulResource}=[resource, text width=7em, fill=yellow!20, draw]
\def\blockdist{2.3}
\def\edgedist{2.5}

\tikzstyle{doc}=[%
draw,
thick,
align=center,
color=black,
shape=document,
minimum width=7mm,
minimum height=13.2mm,
shape=document,
inner sep=2ex,
]

\begin{tikzpicture}
    \node (anchor) [] {};

    \pgfmathsetmacro{\n}{5}
    \foreach \i in {\n,...,1}
    {
        \path (anchor)+(1.0 + 0.1 * \i, -0.1 * \i) node (device\i) [device] {};

        \ifthenelse{\i=1}
        {
          \foreach \j in {\n,...,1}
          {
           \path (device\i.north)+(2.0  + 0.05 * \j, -1.0 - 0.05 * \j) node (computeResource\i_\j) [resource] {Compute Resource \\$c_{0,0} ... c_{0,j}$};
           \path (device\i.north)+(-2.0 + 0.05 * \j, -1.0 - 0.05 * \j) node (memorySpace\i_\j)  [resource] {Memory Space\\ $m_{0,0} ... m_{0,i}$};
          }

          \path (device\i.south)+(0,0.3) node (deviceSubtitle) {Device $d_0 ... d_k$};
        } 
        { } 
    }
    
    \begin{pgfonlayer}{background}
        \path (device1.west |- device1.north)+(-1.2,0.3) node (backgroundA) {};
        \path (device1.south -| device1.east)+(+1.2,-1.0) node (backgroundB) {};
        \path[rounded corners, fill=gray!10, draw=black!50] (backgroundA) rectangle (backgroundB) node (topologyBackground) {};
        \path (device1)+(0.0,-2.00) node (topologySubtitile) {Instance Topology};

        \path (anchor)+(-10.0, 1.95) node (a) {};
        \path (anchor)+(+10.0, 1.95) node (b) {};
        \path [draw, -, dashed, line width=0.5mm] (a) -- (b);
        \path (a) node[align=center,black,anchor=north, yshift=35, rotate=90](managerLayerText) {Managers};
        
        \path (anchor)+(-10.0, -2.75) node (a) {};
        \path (anchor)+(+10.0, -2.75) node (b) {};
        \path [draw, -, dashed, line width=0.5mm] (a) -- (b);
        \path (a) node[align=center,black,anchor=north, yshift=64, rotate=90](managerLayerText) {Stateless};

        \path (anchor)+(-10.0, -5.50) node (a) {};
        \path (anchor)+(+10.0, -5.50) node (b) {};
        \path (a) node[align=center,black,anchor=north, yshift=38, rotate=90](managerLayerText) {Stateful};
        
    \end{pgfonlayer}

    \path (anchor)+(0.25, 3.00) node (MemoryManager) [manager] {Memory Manager};
    \path (MemoryManager)+(+4.00, +0.00) node (TopologyManager) [manager] {Topology Manager};
    \path (MemoryManager)+(-4.00, +0.0) node (CommunicationManager) [manager] {Communication Manager};
    \path (MemoryManager)+(+8.00, +0.0) node (ComputeManager) [manager] {Compute Manager};
    \path (MemoryManager)+(-8.00, +0.0) node (InstanceManager) [manager] {Instance Manager};

     \path (device1.north)+(-8.00, -1.5) node (InstanceTemplate) [resource, text width=7em] {Instance \\ Template};
     
     \path (device1.north)+(+7.5, -1.5) node (ExecutionUnit) [resource, text width=7em] {Execution \\ Unit};

   \path (device1.south)        +(-1.0, -2.5) node (LocalMemorySlot) [statefulResource] {Local \\ Memory Slot};
   \path (LocalMemorySlot.west) +(-2.5,  0.0) node (GlobalMemorySlot) [statefulResource] {Global \\ Memory Slot};
   \path (LocalMemorySlot.east) +(+2.0,  0.0) node (ProcessingUnit) [statefulResource] {Processing \\ Unit};
   \path (ProcessingUnit.east)  +(+3.5,  0.0) node (ExecutionState) [statefulResource] {Execution \\ State};
   \path (GlobalMemorySlot.west)+(-2.5,  0.0) node (Instance) [statefulResource] {Instance};

   \begin{pgfonlayer}{background}

   \path [draw, ->, line width=0.3mm] ([xshift=-30] InstanceManager.south -| Instance.center) --  ([xshift=-30]  Instance.north -| Instance.center) node[align=center,black,anchor=south, xshift=20]() {\textit{detects}};

   \path [draw, -o, line width=0.3mm] ([xshift=20] InstanceManager.south -| Instance.center) --  ([xshift=20]  InstanceTemplate.north -| Instance.center) node[align=center,black,anchor=south, xshift=15]() {\textit{uses}};
   \path [draw, ->, line width=0.3mm] ([xshift=20] InstanceTemplate.south -| Instance.center) --  ([xshift=20]  Instance.north -| Instance.center) node[align=center,black,anchor=south, xshift=20]() {\textit{to create}};

   \path [draw, ->, line width=0.3mm] ([xshift=-10]TopologyManager.south -| TopologyManager.center) --  ([yshift=10, xshift=-10]  device1.north -| TopologyManager.center) node[align=center,black,anchor=south, yshift=-4, xshift=-30]() {\textit{discovers}};
   \end{pgfonlayer}
   
   \path [draw, -o, line width=0.3mm] (MemoryManager.south -| MemoryManager.center) --  (memorySpace1_1.north) node[align=center,black,anchor=south, xshift=30, yshift=-3.0]() {\textit{uses}};
   \path [draw, ->, line width=0.3mm] (memorySpace1_1.south) --  (LocalMemorySlot.north -| memorySpace1_1.center) node[align=center,black,anchor=south, xshift=50, yshift=-3]() {\textit{to allocate / register}};

   \path [draw, -o, line width=0.3mm] ([xshift=-20] ComputeManager.south) --  ([yshift=1.1]computeResource1_1.north) node[align=center,black,anchor=south, xshift=40, yshift=-3.0]() {\textit{uses}};
   \path [draw, ->, line width=0.3mm] (computeResource1_1.south) --  (ProcessingUnit.north -| computeResource1_1.center) node[align=center,black,anchor=south, xshift=25]() {\textit{to create}};

   \path [draw, -o, line width=0.3mm] ([xshift=-10]ComputeManager.south -| ExecutionUnit.center) --  (ExecutionUnit.north -| ComputeManager.center) node[align=center,black,anchor=south, xshift=15, yshift=-1.5]() {\textit{uses}};
   \path [draw, ->, line width=0.3mm] (ExecutionUnit.south -| ComputeManager.center) --  (ExecutionState.north -| ComputeManager.center) node[align=center,black,anchor=south, xshift=25]() {\textit{to create}};

   \path [draw, ->, line width=0.3mm] (CommunicationManager.south -| GlobalMemorySlot.north) --  (GlobalMemorySlot.north) node[align=center,black,anchor=south, xshift=25, yshift=-1.5]() {\textit{exchanges}};

   \path [draw, ->, line width=0.3mm] ([xshift=+20pt]GlobalMemorySlot.south) to[out=-90, in=-90, distance=6mm]  (LocalMemorySlot.south) node[align=center,black,anchor=south, xshift=-40pt, yshift=-30pt]() {\textit{may contain}};

   \path [draw, ->, line width=0.3mm] ([xshift=-20pt]GlobalMemorySlot.south) to[out=-90, in=-90, distance=6mm]  (Instance.south) node[align=center,black,anchor=south, xshift=+40pt, yshift=-30pt]() {\textit{belongs to}};

   \path [draw, ->, line width=0.3mm] (ProcessingUnit.east) --  (ExecutionState.west) node[align=center,black,anchor=center, xshift=-25, yshift=-8pt]() {\textit{executes}};
   
   \path [draw, ->, line width=0.3mm] (ExecutionState.south) to[out=-90, in=-90, distance=6mm] (ProcessingUnit.south)  node[align=center,black,anchor=south, xshift=80pt, yshift=-30pt]() {\textit{yields back to}};

   \path [draw, -, dashed, line width=0.2mm] (InstanceTemplate.north -| InstanceTemplate.east) -- (backgroundA.south);
   \path [draw, -, dashed, line width=0.2mm] (InstanceTemplate.south -| InstanceTemplate.east) -- (backgroundB.north -| backgroundA.center);
   
\end{tikzpicture}
    \end{adjustbox}
\end{center}
\vspace{-10pt}
\caption{Diagram showing the components of the HiCR model and the available operations between them. The model is divided in three component groups: \textit{Managers}, components whose operations represent an application's semantic building blocks; \textit{Stateless}, components that represent static information, and \textit{Stateful}, components with an internal state that mutates over time. }
\label{fig:hicr_model}
\end{figure*}

\cref{fig:hicr_model} shows HiCR's components and the operations defined among them. The model components are divided into three groups: \textit{managers}, \textit{stateless} and, \textit{stateful}. 

Managers are components whose operations have an effect on the system. For example, they can trigger computation operations, the copying of data from one device to another, or create a new application instance. In addition, only managers can create instances of other components, both stateless and stateful.

Stateless components represent information about the system or the static description of a function. As such, these components can be copied, replicated, serialized, and transmitted as required. 

Stateful components represent objects with a finite lifetime whose internal state is subject to change. For example, a running thread or a GPU stream are stateful objects that may be, at any point in the application's time, executing, suspended, or finalized. These components are unique and therefore cannot be replicated.

\subsubsection{Instance Management}
\label{sub:instance_man}

The model defines an \textit{Instance} as any subset of the entire distributed system's available hardware elements, capable of executing independently. An instance is typically implemented as an OS process with full or partial access to a node's CPU, memory, and accelerator devices. Full access is typical for bare-metal deployments, while partial access is typical for virtualized, e.g., cloud-based deployments. The model requires that no two running instances share access to the same devices. Being disjoint, the only contact point for any two instances is via distributed memory communication.

All operations involving instances are handled by the \textit{Instance Manager}. The user may use one of, or a combination of, two ways in which the instance manager enables distributing execution. The first is to detect already created instances. This is typically in cases where the underlying library (e.g., MPI) initiates all instances at launch-time and the instance manager allows retrieving them as a list. The second way is to create new instances during runtime, which is typical for applications that deploy on cloud infrastructures. In this case, the instance manager, running on the initial instance, requests (e.g., to a cloud-service provider) the ramp up of new hosts. Creating a new instance requires passing an \textit{Instance Template} object to the instance manager. This object encapsulates the description of a required topology plus any custom metadata accepted by the underlying technology. This template prescribes the minimal hardware resources required from the new instance. 

Each running instance is semantically equivalent to every other, although only one of them is considered a root instance. A \textit{Root Instance} is either the first instance to be created, or one within the first group of instances created at launch time. The sole purpose of designating an instance as root is to provide a tie-breaking mechanism.

\subsubsection{Topology Management}
\label{sub:topology_man}

A \textit{Topology} represents a full or partial information of an instance's available hardware devices. It comprises a set of \textit{Devices}, a representation of a single hardware element (e.g., a NUMA Domain or a GPU), containing zero or more memory spaces and compute resources.

A \textit{Memory Space} represents a hardware element that exposes explicitly addressable memory segments of non-zero size. Since memory spaces are meant to inform about a device's real memory capacity, the actual physical size is given, and not the size of the virtually addressable space. For example, the system main memory may be exposed as either a single uniform memory access (UMA) memory space (e.g., 128GB), or as multiple non-uniform (NUMA) memory spaces (e.g., 2 x 64GB). For accelerator devices, memory spaces may include device RAM, addressable caches, as well as high-bandwidth memories. Future or non-standard memories, such as large capacity or sequential access, could be represented as long as they fit the definition here provided.

A \textit{Compute Resource} represents a hardware or logical element, capable of performing computation. Typical examples of compute resources are CPU cores, each capable of executing a function independently. They can also represent vector and cube cores in an accelerator, capable of executing discrete kernels and streams. This component contains all the information necessary to uniquely identify the corresponding processor. 

Topologies are discovered by a \textit{Topology Manager}. A combination of different topology managers, each targeting a specific technology, can be used to gather the full information of all the devices comprising the local instance. This information can be serialized and broadcast, allowing users to build a topological picture of the entire distributed system.

\subsubsection{Memory Management}
\label{sub:memory_man}

Memory management in HiCR consists of the creation, exchange and destruction of \textit{Local Memory Slots}. These slots represent the source and destination buffers in data transfers within the scope of a single HiCR instance. They contain the minimum information required to describe a segment of memory (e.g., size, starting address).

The \textit{Memory Manager} object exposes a similar interface to that of the standard C library (i.e., \textit{malloc} and \textit{free}) for the allocation and freeing of local memory slots. However, while the \texttt{std::malloc} operation defaults to obtaining the allocation from the system's main memory, the HiCR model expands on it, allowing the specification of which memory space (and thus, the device) to use as source for the allocation. As long as the memory manager recognizes the specified memory space as one in which it can operate and there is enough space in it, the operation will be successful. 

The model also allows the manual registration of an existing memory allocation as a new local memory slot by specifying its address, size, and the memory space to which it belongs. The memory manager will record the provided information and return a memory slot object that can be used for data transfers. This feature is useful for cases when the user needs to perform a HiCR operation, such as remote communication, on an existing allocation received externally (e.g., from a math library). 

\subsubsection{Communication Management}
\label{sub:comm_man}

All communication in the model is mediated by the \textit{Communication Manager} via its \texttt{memcpy} operation. This operation requires the user to specify the source and destination memory slots, as well as the offsets within them and the size of data to communicate. If the communication manager supports communication between the memory spaces to which each of the memory slots belongs, the operation will be initiated. Otherwise, it will be rejected.

The completion of a memory transfer is not guaranteed after the function call returns. Instead, the communication manager exposes a \textit{fence} mechanism that enables the user to suspend execution until the expected number of incoming and outgoing data transfers have been completed.

The communication manager also is in charge of creating and exchanging \textit{Global Memory Slots}, local memory slots that are made accessible to other HiCR instances and can be used as the source or destination of distributed \texttt{memcpy} operations. The exchange operation communicates the necessary metadata for remote instances to reach the associated memory slot. 

The exchange of global memory slots is a collective operation: all instances must participate by volunteering zero or more local memory slots. The operation returns as many global memory slots as the total amount of local memory slots exchanged. Each of the resulting global memory slots are uniquely identified by a \texttt{tag} and \texttt{key} pair, as defined by the user. The \texttt{tag} element allows for the differentiation of memory slots communicated in different exchange operations, while the \texttt{key} element distinguishes the resulting global memory slots. 

Only three directions are allowed for the \texttt{memcpy} operation: \textit{Local-to-Local}, \textit{Local-to-Global}, and \textit{Global-to-Local}. The first entails two local memory slots and are typically resolved by regular memcpy (e.g., OpenCL's \texttt{clEnqueueCopyBuffer}) operations. The latter involve transfers between instances where one-sided operations (e.g., \texttt{MPI\_Put} and \texttt{MPI\_Get}, respectively) may be used. The \texttt{fence} operation can be used to check for completion in any of these scenarios. On the other hand, \textit{Global-to-Global} operations are not permitted in the HiCR model as it entails communication between two remote instances, neither of which orchestrates the operation.

The model contemplates the existence of heterogeneous systems where hosts and accelerators employ independent interconnects. For example, a data transfer operation may be initiated between a global memory slot representing an allocation in a remote accelerator and a local memory slot in a local accelerator. If available, the communication manager will use the accelerator-specific network to satisfy the request, instead of channeling the transfer through the host network. 

\subsubsection{Compute Management}
\label{sub:compute_man}

The \textit{Compute Manager} is in charge of carrying out computing operations within the HiCR model. Its primary goals involve managing the lifetime of processing units, prescribing the format of execution units, and overseeing the execution of execution states.

A \textit{Processing Unit} represents a compute resource that has been initialized and is ready to execute. For example, a processing unit representing a CPU core (i.e., a compute resource) may be initialized as a POSIX thread with a one-to-one binding to that core. Similarly, a processing unit may represent a stream context in an accelerator device. Upon initialization, the processing unit will keep track of its internal state; i.e., ready, executing, suspended (if supported), or terminated.

An \textit{Execution Unit} is the static description of a function, i.e., a procedure that takes inputs, process them, and produces an output. Examples of execution units are C++ lambda functions, to be executed by a CPU core, or; pre-compiled  kernels, for accelerators. The semantics of an execution unit are given by the user, following the format prescribed by the compute manager.

An \textit{Execution State} represents the execution lifetime of a particular instance of an execution unit, including all the metadata (e.g., inputs, stack, processor state) required to start, suspend and resume (if supported), and finish its execution. 

To carry out the execution, the user asks a compute manager to create a new execution state, derived from a given execution unit. The user must then assign the execution state to an available processing unit. Upon assignment, the processing unit loads the execution state into the processor (e.g., by performing a context switch), and starts computing it. This computation is carried out asynchronously, allowing the rest of the application to perform other operations simultaneously. The completion of an execution state can be queried either in a blocking or non-blocking fashion and, once the execution reaches an end, the execution state is considered finished and cannot be re-used. 

\section{Implementation}\label{sec:implementation}

\begin{figure}[htb]
\begin{center}
\pgfdeclarelayer{background}
\pgfdeclarelayer{foreground}
\pgfsetlayers{background,main,foreground}

\tikzstyle{resource}=[draw, fill=white, font=\footnotesize, text centered, minimum height=1.5em, minimum width=18.5em]
\tikzstyle{backend}=[draw, fill=white, font=\scriptsize, text centered, minimum height=1.5em]
\tikzstyle{device} = [resource, font=\footnotesize, text width=24em, fill=red!20, minimum height=5.0em, rounded corners]
\tikzstyle{manager} = [resource, font=\footnotesize, text width=7em, fill=green!10, minimum height=2.0em, rounded corners]
\tikzstyle{statefulResource}=[resource, font=\footnotesize, text width=7em, fill=yellow!20, draw]
\def\blockdist{2.3}
\def\edgedist{2.5}

\tikzstyle{doc}=[%
draw,
thick,
align=center,
color=black,
shape=document,
inner sep=2ex,
]

\begin{tikzpicture}
    \node (anchor) [] {};

   \path (anchor) node (hicrCoreAPI) [resource] {HiCR Core API};

   \begin{pgfonlayer}{background}
    \path (hicrCoreAPI.west |- hicrCoreAPI.north)+(-0.00,1.60) node (backgroundA) {};
    \path (hicrCoreAPI.east |- hicrCoreAPI.north)+(+0.00,0.35) node (backgroundB) {};
    \path[rounded corners, fill=gray!10, draw=black!50] (backgroundA) rectangle (backgroundB) node (frontendBackground) {};
    \path (backgroundB -| hicrCoreAPI.center)+(0.0,+0.20) node (frontendsTitle) [anchor=center, font=\footnotesize]{Built-in Frontends};     
   \end{pgfonlayer}
    
   \path (backgroundA)+(0.75, -0.5) node (channelsFrontend) [backend, anchor=west] {Channels};
   \path (channelsFrontend.east)+(0.10, +0.00) node (taskingFrontend) [backend, anchor=west] {Tasking};
   \path (taskingFrontend.east)+(0.10, +0.00) node (rpcEngineFrontend) [backend, anchor=west] {RPCs};
   \path (rpcEngineFrontend.east)+(0.10, +0.00) node (dataObjectStoreFrontend) [backend, anchor=west] {Data Objects};

    \path (backgroundA)+(0.00, +0.55) node (applicationCode) [resource, anchor=west] {Application};

    \path [draw, ->, line width=0.5mm] (applicationCode) -- (backgroundA -| hicrCoreAPI.center);
    \path [draw, ->, line width=0.5mm] ([xshift=-75] applicationCode.south -| hicrCoreAPI.center) -- ([xshift=-75] hicrCoreAPI.north -| hicrCoreAPI.center);
    \path [draw, ->, line width=0.5mm] (backgroundB -| hicrCoreAPI.center) -- (hicrCoreAPI.north -| hicrCoreAPI.center);
    
   \begin{pgfonlayer}{background}
    \path (hicrCoreAPI.west |- hicrCoreAPI.south)+(-0.00,-2.10) node (backgroundA) {};
    \path (hicrCoreAPI.east |- hicrCoreAPI.south)+(+0.00,-0.35) node (backgroundB) {};
    \path[rounded corners, fill=gray!10, draw=black!50] (backgroundA) rectangle (backgroundB) node (backendBackground) {};
    \path (backgroundB -| hicrCoreAPI.center)+(0.0,-0.20) node (backendsTitle) [anchor=center, font=\footnotesize]{Built-in Backends};     
   \end{pgfonlayer}

   \path (backgroundA |- backgroundB)+(0.625, -0.75) node (mpiBackend) [backend, anchor=west, minimum width=2em] {MPI};
   \path (mpiBackend.east)+(0.10, +0.00) node (lpfBackend) [backend, anchor=west, minimum width=2em] {LPF};
   \path (lpfBackend.east)+(0.10, +0.00) node (yuanrongBackend) [backend, anchor=west, minimum width=2em] {YuanRong};
  \path (yuanrongBackend.east)+(0.10, +0.00) node (ascendBackend) [backend, anchor=west, minimum width=2em] {Ascend};
  \path (ascendBackend.east)+(0.10, +0.00) node (openclBackend) [backend, anchor=west, minimum width=2em] {OpenCL};
   
  \path  (backgroundA |- backgroundB)+(+1.15, -1.35) node (boostBackend) [backend, anchor=west, minimum width=2em] {Boost};
   \path (boostBackend.east)+(0.10, +0.00) node (pthreadsBackend) [backend, anchor=west, minimum width=2em] {Pthreads};
   \path (pthreadsBackend.east)+(0.10, +0.00) node (hwlocBackend) [backend, anchor=west, minimum width=2em] {HWLoc};
   \path (hwlocBackend.east)+(0.10, +0.00) node (nosvBackend) [backend, anchor=west, minimum width=2em] {nOS-V};

    \path (backgroundA -| backgroundB)+(0.00, -0.60) node (systemLibraries) [resource, anchor=east] {System Libraries};

    \path [draw, ->, line width=0.5mm] (hicrCoreAPI.south -| hicrCoreAPI.center) -- (backgroundB -| hicrCoreAPI.center);
    \path [draw, ->, line width=0.5mm] (backgroundA -| systemLibraries.center) -- (systemLibraries.north -| systemLibraries.center);
\end{tikzpicture}
\end{center}
\vspace{-10pt}
\caption{The current implementation of the HiCR model. Its components and operations are exposed in a \textit{Core API}, which serves as interface between the user-level applications and the underlying system libraries. The core API is distributed together with a set of built-in \textit{backends}, plugins containing the implementation of subsets of model's components for several popular libraries, and \textit{frontends}, HiCR-based libraries providing common, higher-level functionalities. }
\label{fig:hicrImplementation}
\end{figure}
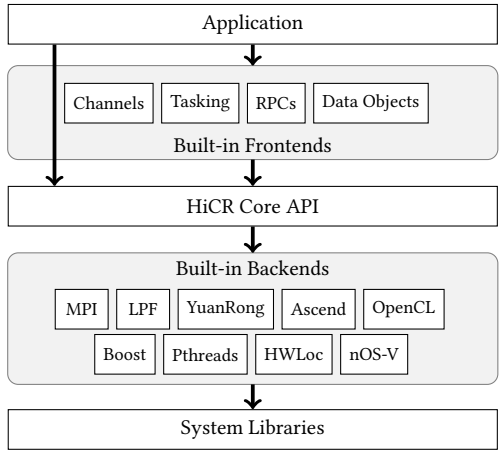

We implemented the HiCR model into an open-source, publicly available C++-based library\footnote{\anonymous{\url{https://github.com/Algebraic-Programming/HiCR}}}. The library is distributed together with a set of built-in backend modules. \textit{Backends} are ready-to-use plugins that translate a subset of the HiCR model into implementation-specific operations that underlying system libraries and device drivers can understand. These details are hidden behind HiCR’s abstract API, allowing a HiCR-based program to preserve its semantics across a diversity of devices. \cref{fig:hicrImplementation} shows how these components serve to relate the user application code to the underlying system libraries.

Built-in backends and their system library dependencies can be included in the compilation chain by configuring them in HiCR's \textit{meson}-based build system during setup time. Any future or third-party backends can be supported by manually adding the corresponding compilation flags. Built-in backends are described in Section~\ref{ssec:builtins}; we first discuss the HiCR interface.


\textit{Frontends} are C++ libraries that contain higher-level out-of-the-box functionality that may be useful for a wide range of applications. The goal of frontends is to facilitate the adoption of HiCR, minimizing the need for users to use its low-level core API directly by providing ready-to-use features such as communication mechanisms, distributed deployment, and topology discovery. Since these libraries are reliant exclusively on calls to the HiCR Core API, they remain implementation-agnostic and their operations can be supported by different backends. An up-to-date list of existing frontends can be found in HiCR's 

\subsection{Programming with HiCR} \label{ssec::programmingHiCR}

All the components of the HiCR model are implemented as C++ abstract classes. As a result, they cannot be instantiated directly. Instead, each of the backends derives them into complete classes by implementing their pure virtual functions. The user is therefore required to first instantiate the appropriate backends and then pass them to a HiCR-based application as inputs.

\cref{fig:example:HiCRBackendsAPI} shows an example of backend instantiation, prior to running a HiCR application. The example starts by initializing the MPI library and passing an MPI communicator object to the MPI instance manager constructor (\cref{code:backendAPI:mpiInit}). Then, it instantiates the \textit{HWLoc} topology and memory managers (Lines \ref{code:backendAPI:hwlocInitStart}-\ref{code:backendAPI:hwlocInitEnd}), which requires passing of an \texttt{hwloc\_topology\_t} object as argument. Finally, it instantiates the \textit{Pthread}-based communication and compute managers (\cref{code:backendAPI:pthreadCommInit,code:backendAPI:pthreadCompInit}). These managers are passed by reference or pointer to a HiCR application, which receives them as abstract classes and thus remains agnostic to the specific choice of backends. 

\begin{figure}[htbp]
\centering
\begin{lstlisting}[style=CppCode,escapechar=|]
// Creating MPI instance manager
MPI_Init(&argc, &argv);
HiCR::MPI::InstanceManager im(MPI_COMM_WORLD); | \label{code:backendAPI:mpiInit} |

// Creating HWLoc topology and memory managers
hwloc_topology_t hwlocObj; | \label{code:backendAPI:hwlocInitStart} |
hwloc_topology_init(&hwlocObj);
HiCR::HWLoc::TopologyManager tm(&hwlocObj);
HiCR::HWLoc::MemoryManager mm(&hwlocObj); | \label{code:backendAPI:hwlocInitEnd} |

// Creating Pthread-based Managers
HiCR::Pthreads::CommunicationManager cmm; | \label{code:backendAPI:pthreadCommInit} |
HiCR::Pthreads::ComputeManager cpm; | \label{code:backendAPI:pthreadCompInit} |
\end{lstlisting}
\caption{Example of backend instantiation. The resulting manager objects are passed to a HiCR application which is built exclusively with calls to abstract HiCR classes.}
\label{fig:example:HiCRBackendsAPI}
\end{figure}

\subsubsection{Example: Inter-Device Communication}

\cref{fig:example:HiCRTbroadcast} shows an example where a given message buffer, stored in a local memory slot, is transferred to all memory spaces at all devices. Since this and the following examples represent pure HiCR applications, all manager objects are pointers to abstract HiCR classes, which are agnostic to implementation decisions. For instance, the code obtains the system's hardware topology from \texttt{tm}, a pointer to the abstract \texttt{HiCR::TopologyManager} class and stores it in a HiCR \texttt{Topology} object (\cref{code:backendAPI:queryTopology}). The communication operations are performed within \texttt{for} loops (\cref{code:coreAPI:deviceForloop,code:coreAPI:spaceForloop}) that iterate among all the memory spaces within all detected devices in the topology and, for each of them, it allocates a new local memory slot (\cref{code:coreAPI:allocate}) and starts a data transfer operation (\cref{code:coreAPI:memcpy}). Finally, it makes sure all the communication operations have terminated (\cref{code:coreAPI:fence}) before returning.

\begin{figure}[htbp]
\centering
\begin{lstlisting}[style=CppCode,escapechar=|]
// Obtain system's CPU topology
HiCR::Topology t = tm->queryTopology(); | \label{code:backendAPI:queryTopology} |

// Broadcast message to all local memory spaces  
for (const auto& d : t.getDevices())  | \label{code:coreAPI:deviceForloop} |
 for (const auto& s : d.getMemorySpaces()) { | \label{code:coreAPI:spaceForloop} |
  auto dst = mm->allocateLocalMemorySlot(s, size); | \label{code:coreAPI:allocate} |
  cmm->memcpy(dst, 0, message, 0, messageSize); | \label{code:coreAPI:memcpy} |
 }
cmm->fence();  // Wait for operations to finish | \label{code:coreAPI:fence} |
\end{lstlisting}
\caption{This example copies a message along all the memory spaces detected by the topology manager. These memory spaces may belong to one or multiple different physical devices on a given node. }
\label{fig:example:HiCRTbroadcast}
\end{figure}

\begin{figure}[htbp]
\centering
\begin{lstlisting}[style=CppCode,escapechar=|]
// Initializing execution in all compute resources
std::vector<HiCR::ProcessingUnit> ps;
for (const auto& d : t.getDevices()) 
 for (const auto& r : d.getComputeResources()) { 
  auto p = cpm->createProcessingUnit(r); | \label{code:exec:puCreate} |
  auto s = cpm->createExecutionState(p, e); | \label{code:exec:sCreate} |
  cpm->initialize(p); | \label{code:exec:executeStart} |
  cpm->execute(p, s); | \label{code:exec:executeEnd} |
  ps.push_back(p);
 }

// Awaiting finalization
for (const auto& p : ps) cpm->await(p); | \label{code:exec:pAwait} |
for (const auto& p : ps) cpm->finalize(p); | \label{code:exec:pFinalize} |
\end{lstlisting}
\caption{This example runs a given execution unit on all of the available compute resources for parallel execution. }
\label{fig:example:HiCRExec}

\end{figure}

\subsubsection{Example: Parallel Execution}

\cref{fig:example:HiCRExec} shows an example where a given execution unit (\texttt{e}) is simultaneously deployed on a set of compute resources. The code first initializes a processing unit for each of the compute resources provided (\cref{code:exec:puCreate}), along with an execution state (\cref{code:exec:sCreate}), and then starts their execution (\cref{code:exec:executeStart,code:exec:executeEnd}). The application then waits for all processing units to finish and terminates the execution states (\cref{code:exec:pAwait,code:exec:pFinalize}), freeing the memory allocated for them.

\subsubsection{Example: Distributed Deployment}

\cref{fig:example:HiCRInstance} shows an example of instance management operations using the HiCR Core API. This example checks whether the \texttt{desired} number of instances have been created at launch time (\cref{code:instanceExample:instanceLaunch}). If not, it tries to create, at runtime, however many of them are found missing (\cref{code:instanceExample:instanceRuntime}). The new instances, if any, are created based on a template that makes sure they satisfy a given set of requirements (\texttt{reqs}), which may include, but are not limited to, hardware or network topology specifications. This snippet is only executed by the root instance (\cref{code:instanceExample:onlyOnce}) to ensure that it runs exactly once.

\begin{figure}[htb]
\centering
\begin{lstlisting}[style=CppCode,escapechar=|]
// If we are not root, return immediately
if (!im->getCurrentInstance().isRoot()) return; | \label{code:instanceExample:onlyOnce} |

// Getting launch-time instances count
const auto instances = im->getInstances();
auto current = instances.size();

// Return if the desired instances already exist
if (current >= desired) return; | \label{code:instanceExample:instanceLaunch} |

// Creating required instances at runtime
auto required = desired - current;
auto temp = im->createInstanceTemplate(reqs);
im->createInstances(required, temp); | \label{code:instanceExample:instanceRuntime} |
\end{lstlisting}
\caption{This example makes sure there are \texttt{desired} number of instances by either having been initially created by an external launcher, or by creating new instances at runtime.}
\label{fig:example:HiCRInstance}
\end{figure}

\subsection{Built-in Backends}\label{ssec:builtins}

\begin{table}[htbp]
    \centering
    \begin{adjustbox}{width=\columnwidth}
    \begin{tabular}{|l|c|c|c|c|c|}
    \hline
        Backend   & Topology & Instance & Communication & Memory & Compute \\ \hline
        MPI       &          & X        & X             & X      &         \\ \hline
        LPF       &          &          & X             & X      &         \\ \hline
        YuanRong  &          & X        &               &        &         \\ \hline
        HWLoc     & X        & X        &               & X      &         \\ \hline
        ACL       & X        &          & X             & X      & X       \\ \hline
        OpenCL    & X        &          & X             & X      & X       \\ \hline
        Pthreads  &          &          & X             &        & X       \\ \hline
        Boost     &          &          &               &        & X       \\ \hline
        nOS-V     &          &          &               &        & X       \\ \hline
    \end{tabular}
    \end{adjustbox}
    \caption{This table lists the backends currently provided and the subset of the HiCR model that they implement.}
 \label{table:implementation:backends}
\end{table}

HiCR can be extended to support any new technologies that satisfy a subset of the core API by developing a new backend plugin for them. Additionally, we distribute with HiCR a collection of ready-to-use built-in backends to support several well-established technologies and devices. \cref{table:implementation:backends} shows the current list of the built-in backends and the manager classes they implement. We chose to provide these backends as they represent a mixture between commonly used technologies with more research-oriented ones.

The \textit{MPI} backend implements operations for communication, memory and instance management using the MPI specification. Its instance manager allows querying how many HiCR instances (i.e., MPI processes) were created at launch time and the unique ID for each of them. The backend instantiates memory slots as MPI \textit{windows} to serve as source and destination for MPI one-sided communication operations and translates distributed HiCR \texttt{memcpy} operations into the corresponding one-sided \texttt{MPI\_Put} or \texttt{MPI\_Get} operations. 

The \textit{LPF} backend provides support for \textit{Lightweight Parallel Foundations} ~\cite{suijlen2019}, a communications library following the BSP model~\cite{valiant1990} for parallel computation. LPF operates mainly by the use of one-sided put and get communication calls whose completion is realized through synchronization calls. This backend uses a synchronization engine implemented on top of the Infiniband Verbs API \cite{infiniband, infinibandVerbsAPI} and provides lightweight synchronization mechanisms based on completion queues.

\textit{YuanRong} \cite{chen2024} is a serverless platform for general-purpose workloads, and tailored for applications running on cloud infrastructures. The serverless paradigm prescribes that applications should be organized into a set of functions that can be dynamically created based on the incoming workload (e.g., numbers of incoming requests). This backend enables launching a HiCR instance as a serverless function at runtime.

The \textit{HWLoc} backend implements topology discovery functionality for CPU-based hosts based on the \textit{Portable Hardware Locality} \cite{hwloc} library. The topology includes a hierarchical view on CPU resources (i.e., sockets, cores, symmetric multi-threading) and their memory and cache structures. It also provides a notion of locality, allowing for determining in which NUMA domain each of the compute resources is located and allocating memory on any such domains.

The \textit{ACL} backend provides support for Huawei NPU accelerators \cite{acl, ascend}, exposing a list of such devices available in the host system and enabling memory allocation both on accelerators' high memory bandwidth (HBM) memory, as well as on the host memory. It allows data motion from either the host to the device and vice-versa, between allocations within the same device, or; between two devices. It also enables the execution of device kernels, building event-based dependency graphs, and querying their progress.

The \textit{OpenCL} backend enables the discovery and use of both CPU and accelerator devices that support the OpenCL \cite{opencl} API. This API offers many of the topology, memory, communication, and computation management support for heterogeneous devices.

The \textit{Pthreads} backend enables threading-based parallelism using the \textit{POSIX Threads} library \cite{posixThreads}. Its compute manager enables the creation of processing units, each one of them representing a system-scheduled thread and mapped 1-to-1 to a CPU core or hyperthread, detected as compute resources by the HWLoc backend. The communication manager employs the standard C \texttt{memcpy} operation, and guarantees correct fencing using mutual exclusion mechanisms. 

The \textit{Boost} backend defines execution units as single (lambda) functions, including the possibility of passing a capture list and a closure argument. It relies on the \textit{Context} library from Boost C++ \cite{boostContext} to instantiate execution units into coroutine-based execution states. These coroutines behave like normal functions, except that they can be suspended and resumed at arbitrary points without the intervention of the OS scheduler.

The \textit{nOS-V} backend enables the use of nOS-V \cite{alvarez2024}, a low-level threading library that enables collaborative co-execution of independent processes. nOS-V features a system-wide scheduler that assigns each task to its own kernel-level thread, all located in a common shared pool across multiple processes. 


\subsection{Built-in Frontends}

Frontends are ready-to-use libraries that expose higher-level features for communication, execution and distributed computing. The ones presented here are fully based on calls to the HiCR Core API, hence preserving the benefits of an implementation-agnostic approach. Here we discuss their rationale and implementation.

The \textit{Channels} frontend is a communication library tailored for frequent and persistent transfer of small data messages across distributed instances. With this frontend, we target applications that operate on a request basis and typically carry a quality-of-service (QoS) requirement of a low-latency result turnover. Channels operate by exchanging pre-allocated circular buffers between the sender and receiver instances. These buffers will serve as destination for the transmitted messages. By pre-allocating the buffers, the producer knows where to push the next message, as long as the buffer has not filled up. At that point the producer may not send any more messages until the consumer notifies that a message has been consumed. This logic enables both ends to decouple transfer and synchronization messages, allowing for both minimal per-message handshaking as well as throughput-oriented channel implementations. 

The library supports both \textit{Single Producer, Single Consumer} (SPSC) and \textit{Multiple Producer, Single Consumer} (MPSC) paradigms. To support multiple producers, the library offers two operating modes: \textit{locking}, where a collective exclusive access guarantees a shared channel does not overflow, and \textit{non-locking}, where dedicated buffers per producer are employed, eliminating the expensive collective exclusive  access, but increasing memory requirements.

The \textit{Data Object} frontend is a communication library, tailored for sporadic communication of large data objects, such as multi-dimensional tensors. This library allows for performing communication operations without the need of pre-exchanged buffers. Instead, it works by creating a \textit{Data Object} which represents a block of data contained inside a local memory slot, and making it remotely accessible to any other instance by a call to a \texttt{publish} operation. Upon publication, the caller obtains a unique data object identifier that can be exchanged with other instances via the Channels frontend. To access a published data object, other instances perform a \texttt{getHandle} operation, which takes a unique identifier as argument and returns a \textit{handle} to the remote object. This handle only represents the required metadata to retrieve the remote object. The data itself can be obtained with a call to \texttt{get}, which takes the handle as argument. This function initializes an asynchronous data transfer whose completion can be fenced later using a mechanism similar as described in Figure~\ref{fig:example:HiCRTbroadcast}. 

The \textit{RPC} frontend offers mechanisms for the registration, listening, and execution of Remote Procedure Calls. This interface is crucial for the initial coordination of execution among multiple instances, especially in the context of applications that rely on instance creation at runtime. RPCs can be used to exchange information about the instance's topology information, establish the creation of communication channels, and to coordinate task execution. To execute an RPC, the function must be pre-registered on the receiving instance. Then, the receiving instance must then enter a listening state either before or after the caller instance launches an RPC request. After execution, the receiving instance may produce a return value that will be automatically returned to the caller.

The \textit{Tasking} frontend contains building blocks to develop a task-based runtime system. It provides basic support for stateful tasks with settable callbacks to notify when the task changes state, e.g., from \textit{executing} to \textit{finished}. It also contains support for stateful worker objects. These objects contain a simple loop that calls a \textit{pull} function, i.e., a user-defined scheduling function that should return the next task to execute (or a \textit{null} pointer, if none is available). The frontend requires specifying two, possibly distinct, compute managers: one for the worker objects and another for the tasks. In this way, the frontend allows, for example, managing scheduling on the CPU, while executing tasks directly on an accelerator device. This frontend also contains an interface to \textit{OVNI} \cite{ovni2025}, an instrumentation library to register execution traces. These traces are collected regardless of the computing backend selected and can be loaded into any performance analysis tool.

\section{Experimental Evaluation} \label{sec:experiments}

Here we present experimental results based on reproducible test cases programmed exclusively with calls to the HiCR API. The goal of these experiments is to demonstrate the adaptability of HiCR-based applications to different execution environments by selecting the appropriate backend for each case, without requiring changes in its source code. The source code required to reproduce these experiments is available in HiCR's public repository.

\subsection{Test Case 1: Communication Benchmark}\label{sec:Test_case_1}
\label{sec:miniapp1}

This test case involves launching two instances communicating through two opposing single-producer single-consumer channels for bi-directional communication. It allocates a short message buffer at the consumer side with a fixed single-message capacity. After sending a message (ping), the sender waits on receiving the echoed message (pong). This exchange forces a ping-pong pattern (similar to a one-sided version of the two-sided \textit{NetPIPE} \cite{snell1996netpipe}), which results in either latency-bound results when communicating small messages, or throughput-bound results for large messages.

\begin{figure}[htbp]
  \centering
  \includegraphics[width=0.5\textwidth]{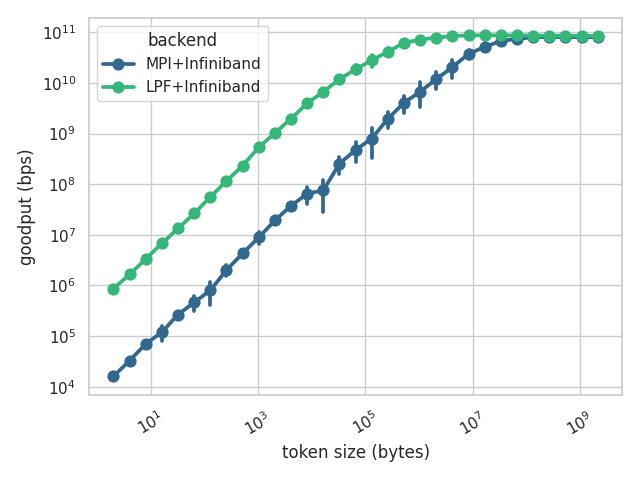}
  \vspace{-8mm}
  \caption{Observed goodput for ping-pong benchmarks using the LPF (top series) and MPI (bottom series) backends over multiple message sizes. LPF relies on IBverbs directly while MPI relies on OpenMPI RMA primitives.}
 \label{fig:experiments:pingpongResult}
\end{figure}

We designed the benchmark to compare the performance of two backends: (a) the LPF backend, relying on its \emph{zero} engine \cite{lpfNoC} that employs primitives for Infiniband networks, and (b) the MPI backend. All tests were conducted on a cluster equipped with a Mellanox EDR 100 Gbps Infiniband fabric. We run the test with a range of message sizes from $1$ byte to $\approx 2.14$ Gigabytes, repeating each test 10 times and showing the standard deviation vertically. The results of the ping-pong benchmarks are shown in ~\cref{fig:experiments:pingpongResult}, measuring the \textit{goodput} $G(s)$ (i.e., effective throughput) across different message sizes.

For small messages, the LPF backend achieves a $~70\times$ increase in goodput compared to the MPI-based one. We attribute this to LPF making direct use of hardware-enabled Infiniband completion queues that minimizes handshaking. This extends beyond what the MPI standard currently allows, similarly to other proposed optimizations~\cite{Belli2015,Sergent2019}. The MPI backend employs standard one-sided MPI primitives that induce less efficient handshaking. The larger messages ($>10^9$ bytes) are far less affected by handshaking latencies, and therefore indeed the goodput of both backends converges to ${\sim}80\%$ of the maximum theoretical throughput of the interconnect: 100 Gbps.


\subsection{Test Case 2: Heterogeneous Inference}

This test case implements a simple forward inference pipeline, mimicking typical AI inference workloads. We implemented a HiCR application featuring a neural network that takes an encoded image as input and outputs the most likely digit (from 0 to 9) that they represent. We used the \textit{MNIST} \cite{mnist} dataset to train the network and saved its weights for later use during the inference. We run the application passing to it three different backends: Pthreads, ACL, and OpenCL. For each backend, we provide an appropriate kernel function: the Pthreads variant runs \textit{OpenBLAS} \cite{openblas} dense linear algebra kernels; the ACL variant uses pre-compiled kernels provided with the NPU device, and; the OpenCL variant uses a na\"ive version of the required kernels. We conducted experiments on two computing nodes: (A) containing an Intel Xeon W-1270 CPU and an Intel P630 GPU, and (B) containing a Huawei Kunpeng 920 CPU and a Huawei 910A NPU. To prove the correctness, we developed an ad-hoc C++ implementation of the test using OpenBLAS kernels directly without the use of HiCR. We used the latter to verify that all variants produce consistent results on each architecture.


\begin{table}[t]
    \centering
    \begin{adjustbox}{width=\columnwidth}
    \begin{tabular}{|l|c|l|c|c|}
    \hline
        Device        & Node & Backend    & Accuracy  & img-0 score           \\
        \hline
        W-1270          & A    & pthreads   & 94.64\%    & 9.921433449            \\
        P630            & A    & opencl     & 94.64\%    & 9.921431541            \\
        Kunpeng 920     & B    & pthreads   & 94.64\%    & 9.921433449            \\
        Huawei 910A     & B    & acl        & 94.64\%    & 9.921875000            \\ 
        \hline
    \end{tabular}
    \end{adjustbox}
    \caption{Inference results. The table reports the percentage of correct predictions over the entire test set (10'000 images), and the highest score computed for img-0 of the test set.}
 \label{table:experiments:heterogeneousResult}
\end{table}

\cref{table:experiments:heterogeneousResult} presents the inference results, showing the prediction accuracy for each architecture and backend, and also the highest score given to the most probable digit in the first image of the set (\textit{img-0}). The table shows that, while all backends show consistent accuracies, they still produce slight variations in precision. These variations can be attributed to differences in the floating-point precision of the devices and to differing orders of computations in the compute kernels. Despite these, this test shows that a given HiCR-based application can be executed equally on either CPU or accelerator devices by providing the appropriate backend and the kernels to run its operations.



\subsection{Test Case 3: Fine-Grained Tasking}\label{sec:Fine-Grained}

This test case computes the Fibonacci number $F(n)$ using a na\"ive approach, i.e., recursively computing $F(n-1)$ and $F(n-2)$ as independent tasks until reaching $F(1)$ and $F(0)$. To schedule the task dependency graph, we developed a HiCR-based lightweight scheduler\footnote{\anonymous{https://github.com/Algebraic-Programming/TaskR}} that assigns tasks to worker threads as they become available. Since we use the exact same setup for all runs, this test is designed to compare the compute performance between different HiCR backends in a way that measures the overheads of context switching.

We employ two HiCR variants: (a) one using the nOS-V threading backend for both worker and tasks management, and; (b) another using Pthreads + Boost for thread-based workers and coroutine-based tasks, respectively. For both variants, we compute $F(24)$ with an expected result of $46\ 368$, requiring the execution of $150\ 049$ tasks in total. We ran benchmarks on a dual-socket $22$-core Intel Xeon Gold 6238T server with hyperthreading enabled, and report the best measured time among $10$ runs. We determined that using $8$ worker threads that are pinned to individual cores in the same socket yields the best performance for both variants.

\begin{figure}[t]
  \centering
  \includegraphics[width=\columnwidth]{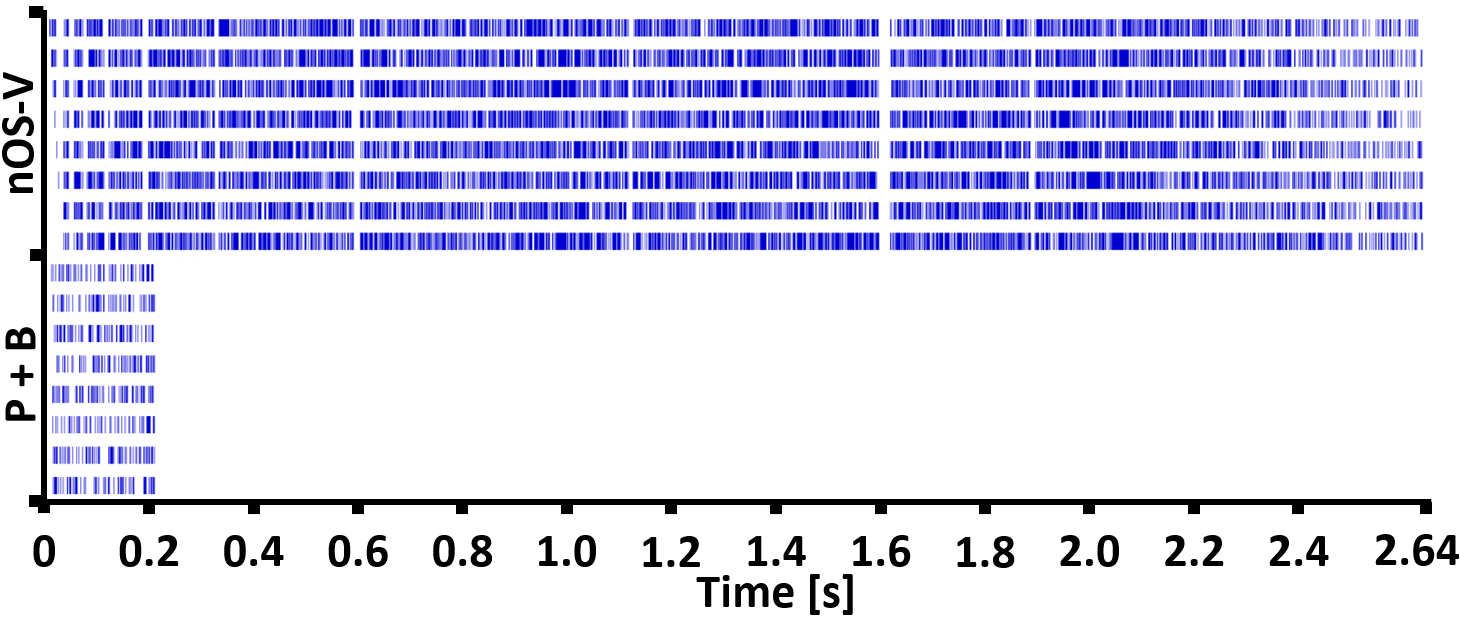}
  \vspace{-5mm}
  \caption{Execution timelines for the Fibonacci example that executes $150\ 049$ tasks using $8$ cores. Each horizontal line represents the timeline of a CPU core, with solid traces indicating meaningful work and empty spaces indicating scheduling overhead.}
 \label{fig:experiments:fibonacci}
\centering
  \includegraphics[width=\columnwidth]{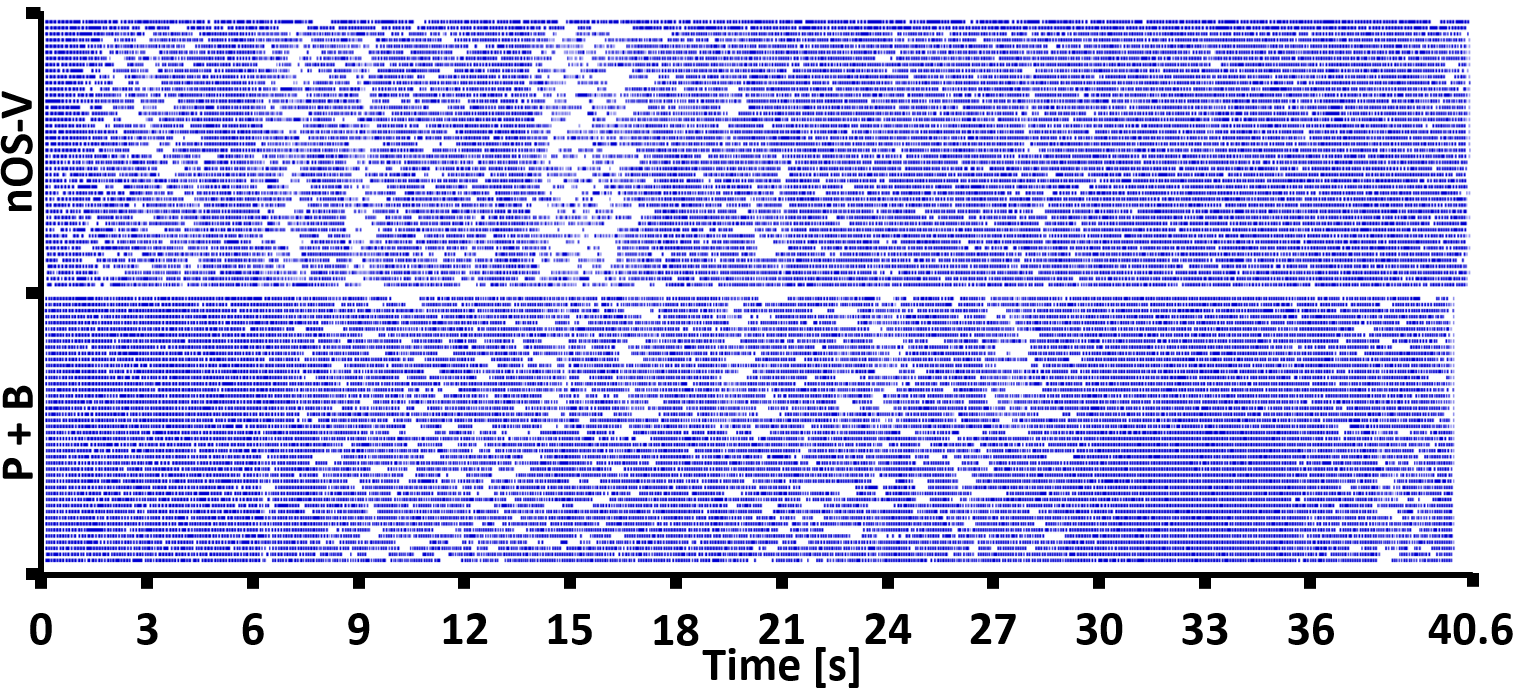}
  \vspace{-5mm}
  \caption{Execution timelines for the Jacobi example running 500 iterations using $1\times2\times22=44$ threads of a Intel Xeon Gold 6238T system with hyperthreading enabled (but here unused).}
 \label{fig:experiments:jacobi}
\end{figure}

We used the OVNI library to obtain the core execution traces and visualize them with \textit{Paraver} \cite{paraver2025}. \cref{fig:experiments:fibonacci} shows the traces for the best result for each variant, where nOS-V and Pthreads + Boost backends finish execution in 1.34 and 0.21 seconds, respectively. This example demonstrates that user-level context switching between fine-grained tasks can greatly reduce overheads compared to delegating scheduling decisions to the OS.

\subsection{Test Case 4: Coarse-Grained Tasking}\label{sec:Coarse-Grained}



Consider a three-dimensional iterative heat equation solver that uses the Jacobi method and a $13$-point averaging stencil with Manhattan distance one and a finite element grid of size $704^3$ elements. Our test case divides the grid, consisting of a single contiguous allocation, across $l_x \times l_y \times l_z$ local subgrids, and assigns each to a unique worker thread. Each thread thus executes an expensive local computation that dominates the total runtime, followed by communication of the subgrid halos. This process repeats for a set number of iterations.

We compare the nOS-V and Pthreads+Boost variants over $500$ iterations of the solver and report the best measured time among $20$ runs. For this case, we employ the same dual-socket system as in Section~\ref{sec:Fine-Grained} and achieve best performance using a $1\times2\times22$ thread grid, thus spawning a total of $44$ threads that are each assigned a core; i.e., not using hyperthreading.
 \cref{fig:experiments:jacobi} shows the resulting traces for the best run for each variant, where the nOS-V finishes execution in $40.5$ s ($43.1$ GFlop/s), and for Pthreads + Boost, in $39.9$ s ($43.7$ GFlop/s). In this case, we used HiCR to show that the effects of scheduling overheads are minimal when running coarse-grained tasks, regardless of the choice of backend. The benefits of nOS-V may hence be exploited without noticeable performance impact.

\begin{figure}[t]
  \centering
  \includegraphics[width=\columnwidth]{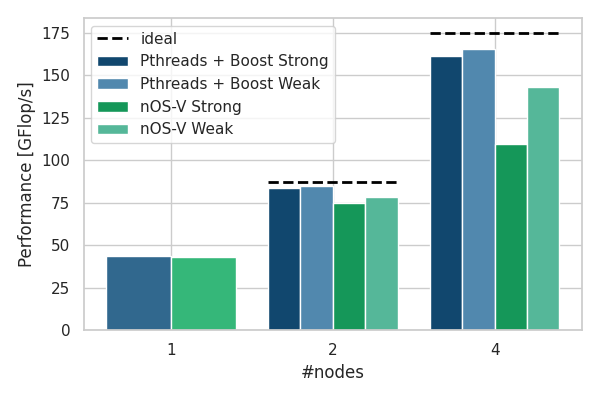}
  \vspace{-5mm}
  \caption{Strong and weak scaling of the Jacobi example using up to $4$ nodes of dual-socket Intel Xeon Gold 6238T CPUs. Each node has $22$ cores per socket with hyperthreading enabled (but here unused).}
 \label{fig:experiments:strong_weak}
\end{figure}

We also evaluate the impact of the computing backend choice on scaling and communication overheads on up to $4$ nodes, where each node consists of the same dual-socket system previously used. To run this test, we scale the number of HiCR instances with the number of nodes where each instance handles one rectangular cuboid resulting from splitting the mesh in $p_x  \times  p_y  \times  p_z$ parts, where $p_xp_yp_z=p$ the total number of nodes used. After each iteration, halos are exchanged via distributed-memory communication through the LPF Infiniband backend from Section~\ref{sec:Test_case_1}. We observe the best performance using a $p\times1\times1$ node mesh, which, combined with the thread mesh from the previous experiment, results in a logical $p\times2\times22$ mesh. \cref{fig:experiments:strong_weak} shows the resulting weak and strong scaling for both variants. For weak scaling, we increase the number of elements to $880^3$ for $p=2$, and to $1056^3$ for $p=4$. 

Pthreads+Boost consistently achieves a better performance than nOS-V in all cases, which, after analysis, we attribute to the use of nOS-V resulting in eager polling of the completion status of distributed-memory communication. This, in turn, causes threads interfering with one another during the communication phase.

\section{Conclusions and Future Work}\label{sec:conclusion}

The latest developments in distributed heterogeneous technologies present new challenges for programmers. Developing applications in a way that they can adapt to and stay current with the latest and upcoming hardware can be a daunting endeavor. We have presented a model to minimize this effort, enabling the expression of program semantics that are independent from specific underlying system technologies, bypassing the complexities of writing technology-specific code. Contrary to other solutions, HiCR describes a set of abstract operations without prescribing particular programming models or paradigms, and without making any other implementation decisions. These facets make HiCR a suitable model for a thin Runtime Support Layer that resides between DSLs, libraries, and programming frameworks on the one hand, and raw system technologies on the other.

We have shown empirically that HiCR programs can execute on multiple architectures and employ different supporting libraries, simply by switching out the underlying HiCR backends-- and that doing so preserves the overall program semantics. While these experiments also show that applications may be implemented directly on top of the HiCR model, we posit its use as a Runtime Support Layer is of higher value.


Future work includes extending the model for discovery of the interconnect topology, associating latency and bandwidth capabilities to both memory spaces (e.g., in NUMA systems) and interconnect links~\cite{suijlen2019}, and the inclusion of (distributed) file management, multi-user job allocation, fault tolerance, and security isolation.

\bibliography{bibliography}

\end{document}